\documentclass[prl,aps,twocolumn,showpacs,amsmath,amssymb, superscriptaddress]{revtex4-1}

\usepackage{graphicx}
\usepackage{bm}

\begin{document}

\title{Finite size behaviors of critical Ising model on a rectangle with free boundaries}
\author{Xintian Wu}
\email{wuxt@bnu.edu.cn}
\affiliation{Department of Physics, Beijing Normal University,
Beijing, 100875, China}

\author{Nickolay Izmailian}
\email{izmail@yerphi.am}
\affiliation{Department of Physics, Beijing Normal University,
Beijing, 100875, China} \affiliation{A.I. Alikhanyan National
Science Laboratory, Alikhanian Br.2, 375036 Yerevan, Armenia.  }
\author{ Wenan  Guo }
\email{waguo@bnu.edu.cn}
\affiliation{Department of Physics, Beijing Normal University,
Beijing, 100875, China}

\date{\today}

\begin{abstract}

Using the bond-propagation algorithm, we study the 
Ising model on a rectangle of size $M \times N$ with free
boundaries. For five aspect ratios $\rho=M/N=1,2,4,8,16$, the
critical free energy, internal energy and specific heat are
calculated. The largest size reached is $M \times N=64\times 10^6$. The accuracy of the free energy reaches $10^{-26}$. Basing on these accurate data, we determine exact expansions of the
critical free energy, internal energy and specific heat. With these expansions, we extract the bulk, surface and corner parts of free energy, internal energy and specific heat.
The fitted bulk free energy density is given by $f_{\infty}=0.92969539834161021499(1)$, comparing with Onsager's exact result $f_{\infty}=0.92969539834161021506\cdots$.  We prove the conformal field theory(CFT) prediction of the corner free energy, in which the central charge of the Ising model is found to be $c=0.5\pm 1\times 10^{-10}$ comparing with the
CFT result $c=0.5$. We find that not only the corner free energy but also the corner internal energy and specific heat are geometry independent, i.e., independent of aspect ratio. The implication of this finding on the finite scaling is discussed.
In the second order correction of the free energy, we prove the geometry dependence predicted by
CFT and find out a geometry independent constant beyond CFT. High order corrections are also obtained.
\end{abstract}

\pacs{75.10.Nr,02.70.-c, 05.50.+q, 75.10.Hk}

\maketitle

{\it Introduction}. Finite size effect has been attracting
tremendous interest  in the study of condensed matter physics. It
becomes of practical interest due to the recent progresses in fine
processing technologies, which has enabled the fabrication of
nanoscale materials with novel shapes \cite{kawata,puntes,yin}.
Exact solutions have been playing a key role in determining the
form of finite size scaling. Ferdinand and Fisher
\cite{fisher1969} pioneered on the 2D Ising model on a finite size
lattice, which extended Onsager's exact solution \cite{onsager}
and stimulated the ideas of finite size scaling. Since then, exact
results of the model on finite size lattices with various
boundaries have been studied intensively
\cite{onsager,kaufman,fisher1969,fisher,izmailian2002a,izmailian2002b,izmailian2007,Salas,Janke}.
Detailed knowledge has been obtained for the torus case
\cite{izmailian2002a,Salas}, for helical boundary condition
\cite{izmailian2007}, for Brascamp-Kunz boundary condition
\cite{izmailian2002b,Janke} and for infinitely long cylinder
\cite{izmailian1}. However in the jigsaw puzzle of the 2D Ising
model an important piece  is still missed, which is the exact
result on a rectangle with free boundaries. Generally speaking the
rectangle geometry system is very interesting in its own right. It
is, for instance, the natural one to consider in the case of
quenches for one dimensional quantum systems with open boundaries.
In 2D it is the simplest geometry to study transport properties
for Anderson localization. Although there are Monte Carlo and
transfer matrix studies on this problem \cite{landau, stosic}, the
accuracy or the system sizes of the results are not enough to
extract the finite size corrections. Meanwhile, for 2D  critical
systems, a huge amount of knowledge has been obtained by the
application of the powerful techniques of integrability and
conformal field theory (CFT) \cite{blote, cardy, kleban}. Cardy
and Peschel predicted that the next subdominant contribution to
the free energy on a square comes from the corners \cite{cardy},
which is universal, and related to the central charge $c$ in the
continuum limit. Kleban and Vassileva \cite{kleban} extended the
study of the free energy on a rectangle. They further derived  a
geometry dependent term to the free energy. However, they did not
determine a geometry independent additive constant in the
coefficient. Till now there is few evidence  for these predictions
from exact solutions or numerical calculations. Furthermore, as
far as we know, there is no detailed study of the internal energy
and the specific heat on a rectangle neither by CFT nor by exact
solutions/numerical calculations.

Recently  an efficient bond propagation (BP) algorithm was
developed for computing the partition function of the Ising model
in two dimensions, which is exact to machine precision and
works for any planar network of Ising spins with arbitrary bond
strengths \cite{loh1,loh2}.
It is also much faster than Monte Carlo simulation, and costs quite
moderate memory comparing with the transfer matrix method.
Very large system size can be reached.
The BP algorithm is thus a powerful tool to study
the Ising model on a rectangle with free edges and corners. In
this letter we apply the BP algorithm to study the Ising model on an
$M\times N$ rectangle. We obtain finite size data of the critical free energy $f$, internal
energy $U$ and specific heat $c$. By fitting these data we find that
the exact expansion of the critical free energy, internal
energy and specific heat can be written in the following form
\begin{equation}
f=f_{\infty}+f_{surf}\frac{M+N}{S}+f_{corn}\frac{\ln
S}{S}+\sum_{k=1}^{\infty}\frac{A_{k}}{S^{(k+1)/2}}, \label{freeenergy}
\end{equation}
\begin{equation}
U=U_{\infty}+U_{surf}\frac{M \ln N+N\ln M}{S} +U_{corn} \frac{\ln
S }{S} +\sum_{k=1}^{\infty}\frac{B_k}{S^{k/2}},
\label{internalenergy}
\end{equation}
\begin{eqnarray}
c&=&A_0\ln N+c_0+c_{surf} \frac{M\ln N+N\ln M}{S} +c_{corn}\frac{\ln
S}{S} \nonumber \\
&&+\sum_{k=1}^{\infty}\frac{D_k}{S^{k/2}}, \label{specificheat}
\end{eqnarray}
where $S=M\times N$ is the area of the system. $f_{\infty},
U_{\infty}$ are the bulk term, $f_{surf}$, $U_{surf}$ the surface
coefficient, $f_{corn}, U_{corn}$ the corner coefficient for the
free energy, internal energy respectively. $c_{surf}, c_{corn}$
are the corresponding coefficients for the specific heat. We find
the fitted values $f_{\infty},f_{surf},U_{\infty},U_{surf},A_0$
are excellently consistent to the exactly known results
\cite{onsager,kaufman,fisher1969,fisher,izmailian2002a,izmailian2002b,izmailian2007}.
The corner free energy $f_{corn}=c/8$ \cite{cardy, kleban} is proved.
More over 
the geometry independent constant, which is ignored in Kleban and
Vassileva work \cite{kleban}, is determined. We also find the
corner contributions $U_{corn}$ and $c_{corn}$, which are
independent of aspect ratio. As far as we know, no previous
studies predict such terms. We start by introducing the BP
algorithm briefly. Then we present our numerical results and
analysis.

\begin{figure}
\includegraphics[width=0.5\textwidth]{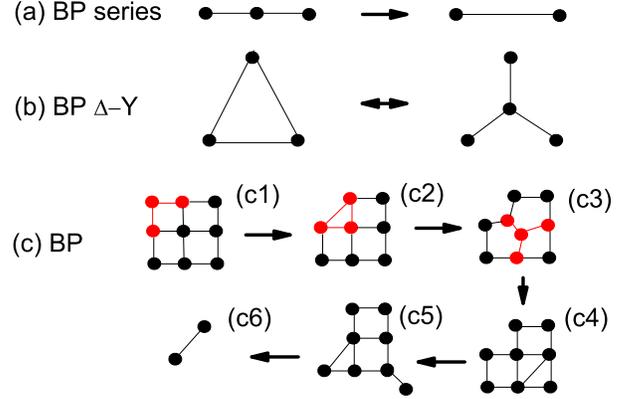}
\caption{(Color online) (a) and (b) are building blocks of the BP algorithm: BP series and BP $\Delta -Y$ transformation, respectively.
  (c) The schematic of BP algorithm. (c1) The BP series is applied to the three spins and two bonds in red.
  (c2) The $\Delta -Y$ transformation is applied to the three spins and three bonds in red.
  (c6) Finally only two spins and one bond are left.}
\label{bp}
\end{figure}

{\it Method}. The schematic of the BP algorithm is shown in Fig.
\ref{bp}. In this algorithm BP series reduction, BP Y-$\Delta$
transformation and its inverse are the building blocks. By
successively integrating in and then integrating out spin degrees
of freedom in a way that only introduces local changes to the
network, this algorithm progressively moves degrees of freedom to
an open edge of the network, where they are eliminated. The
transformation in each step is exact. The numerical accuracy is
limited by machine's precision, which is the round-off error
$10^{-32}$ in the quadruple precision. The BP algorithm needs
about $N^3$ steps to calculate the free energy of an $N\times N$
lattice
(much faster than other numerical method).
Therefore the total error is approximately
$N^{3/2}\times 10^{-32}$. This estimation has been verified in the
following way: We compared the results obtained using double
precision, in which there are 16 effective decimal digits, and
those using quadruple precision. Because the latter results are
much more accurate than the formal, we can estimate the error in
double precision results by taking the quadruple results as the
exact results. We thus found that the error is about $N^{3/2}
\times 10^{-16}$. In our calculation, the largest size reached is
$M=N=8000$, the round-off error is less than $10^{-26}$.
The partition function of the Ising model on a 2D square lattice
is
\begin{equation}
Z=\sum_{\{\sigma_i\}}\exp{(\beta \sum_{<i,j>}\sigma_i\sigma_j)} \label{eq:partition},
\end{equation}
where the nearest neighbor couplings are dimensionless and $\beta$ is the
inverse temperature.
The free energy density, internal
energy per spin and specific heat density are defined by
\begin{equation}
f=\frac{\ln Z}{M N}, \hskip 0.5cm U=\frac{\partial f}{\partial
\beta}, \hskip 0.5cm c=\beta^2 \frac{\partial^2 f}{\partial
\beta^2},
\end{equation}
respectively.
With the BP algorithm, we get the free energy density $f$ at the critical 
$\beta_c=\frac{1}{2}\ln(1+\sqrt{2})=0.44068679\cdots$
directly. 
The internal energy and specific heat are calculated by using a
differentiation method
\begin{eqnarray}
U & \approx & -\frac{f(\beta_c+\Delta \beta)-f(\beta_c-\Delta
\beta)}{2\Delta \beta} ,\nonumber \\
 c & \approx & \beta_c^2\frac{f(\beta_c+\Delta \beta)+f(\beta_c-\Delta
\beta)-2f(\beta_c)}{(\Delta \beta)^2}.
\end{eqnarray}
In our calculation, $\Delta \beta=10^{-7}$ is used.
The error in the
calculation of $U$ due to the finite $\Delta \beta$ is approximately
$\frac{1}{3}(\Delta \beta)^2 \partial^3 f /
\partial \beta^3 $, which is less than $ 10^{-11}$
for the lattice with $N\le 2000$.
The error in
the calculation of $c$ due to finite $\Delta \beta$ is
$\frac{1}{12}(\Delta \beta)^2
\partial^4 f / \partial \beta^4 $, which is less than $10^{-9}$ for
$N\le 2000$.
Another source of error is the accumulated error due to round-off,
which is less than $10^{-26}$. This error is
amplified by $1/\Delta \beta(1/{\Delta \beta}^2)$ times in the
calculations of $U$ ($c$), 
which is thus around $10^{-19}$ ($10^{-12}$).
Therefore the errors of $U, c$ are mainly caused by the finite
$\Delta \beta$ in the differentiation. In other words the
accuracies of the free energy, internal energy and specific heat are
$10^{-26},10^{-11},10^{-9}$ respectively.

The calculations have been carried out for various aspect ratios
$\rho=M/N=1,2,4,8,16$ on an $M\times N$ rectangular lattice. For
$M=N$, the calculation was carried out from $N=30$ to $N=8000$.
For $\rho=2,4,8,16$, the calculated lattice sizes are from $N=30$
to $N=2000$.

{\it Critical free energy}.  We fit the data of free energy
density with the formula given by Eq. (\ref{freeenergy}) 
with $k$ from 1 to 8.
The fitting method is the Levenberg-Marquardt method \cite{LM} for nonlinear
fit.
The standard deviation (SD) is defined by
$SD=\sqrt{\sum_i(f_i-f_i^{(fit)})^2/(n_d-n_f)}$ with $f_i$
the numerical data, $f_i^{(fit)}$  the value given by the
fitting formula, $n_d$  the data number and $n_f$ the number of
fitting parameters. 
For all $\rho$, SD reaches $10^{-20}$.

The high accuracy can be seen from the bulk value $f_{\infty}$.
For $\rho=1$, the fitted values of $f_{\infty}$ is
$0.92969539834161021499(1)$. The asymptotic bulk value of the free
energy density should be the same as the exact result given by
Onsager\cite{onsager}, i.e., $f_{\infty}=\ln
\sqrt{2}+\frac{2}{\pi}G=0.92969539834161021506\cdots$, where
$G=1-\frac{1}{3^2}+\frac{1}{5^2}-\frac{1}{7^2}+\cdots$.
Our estimation is consistent to it in the accuracy of $10^{-20}$.
For other $\rho$ the consistency is the same.

According to finite size scaling, the surface correction term 
stems from free edges.
This correction for the model on an infinitely
long strip with two free edges has been found through exact solution to be
$(D_1-\frac{1}{2}\ln(1+\sqrt{2}))/N$, where
$D_1=\int_{0}^{\pi}\ln[1+\sqrt{2}(1-\cos
\theta)^{1/2}(3-\cos\theta)^{-1/2}]\approx 0.2589553765253$
\cite{fisher}. Note the infinitely long strip can be considered as
a rectangle with a aspect ratio $\rho \to \infty$. Thus
the surface correction for a rectangle should be $f_{surf}
(M+N)/S$ with $f_{surf}=D_1-\frac{1}{2}\ln(1+\sqrt{2})=-0.1817314169844187569
\cdots$.
Our fitting results coincide with this prediction precisely.
For $\rho=1,2$, the fitted values  $-f_{surf}$ are
$0.18173141698441877(3)$, $0.1817314169844188(1)$ respectively,
and for $\rho=4,8,16$, they are all $0.1817314169844188(2)$. The
correction term $f_{corn} \ln S/S$ stems from the corners of the
rectangle.  According to  Cardy and Peschel \cite{cardy}, the four
corners, at $M=N$, give rise to the term $\frac{c }{4 }\ln
(N)/N^2$, where $c=1/2$ is the central charge.
Kleban and Vassileva \cite{kleban} extended the results to 
a rectangle with $M\neq N$ using CFT: The
corner free energy is $c \ln (S)/(8S)$, which yields
\begin{equation}
f_{corn} =\frac{c}{8}=0.0625~. \label{eq:f-corn}
\end{equation}
In our fitting  $f_{corn}$ are found to be $0.0625+(2 \pm 3)\times
10^{-14}$, $0.0625+(1 \pm 2)\times 10^{-12}$,$0.0625+(1 \pm
5)\times 10^{-12}$,$0.0625+(1 \pm 5)\times 10^{-12}$,$0.0625 \pm
1\times 10^{-11}$ for $\rho=1,2,4,8,16$ respectively. All of them
lead to the central charge $c=0.5$ with the error less than $
10^{-10}$.
There is a $1/N^2$ correction for the infinitely long strip
\cite{fisher, blote} with the coefficient
$c \pi/24 \approx 0.0654498 $.
Kleban and Vassileva \cite{kleban} proved that, for a finite rectangle,
the correction is proportional to $1/S=1/(M\times N)$ with
the coefficient
\begin{equation}
A'_1=-\frac{c}{4}\ln[\eta(q)\eta(q')] , \label{eq:kleban}
\end{equation}
where $\eta(q)=q^{1/24}\prod_{n=1}^{\infty}(1-q^n)$ with
$q=e^{-2\pi \rho}, q'=e^{-2 \pi/\rho}$. At the limit $\rho \to
\infty$, $A'_1 \rho^{-1} \to c \pi/24$ which recovers the
infinitely long strip result.

However, Kleban and Vassileva mentioned that, in their derivation,
a possible geometry-independent additive constant was ignored
\cite{kleban}. In other words, the coefficient of  $1/S$ should be
\begin{equation}
A_1=A'_1+F_0,
\end{equation}
where $F_0$ is the constant which contributes $F_0/S=F_0/(\rho N^2)$
to the free energy density, which tends to 0 in the infinitely long strip limit.
For $\rho=1,2,4,8,16$, the values of $A'_1$ are $0.065918017562$,
$0.087578866954$, $0.175155990232$, $0.393633679243$,
$0.873910756056$ respectively.
 By comparing the fitted $A_1$ (see
Tab. \ref{fitfen}) and the theoretical $A_1'$ , we obtained this
constant
\begin{equation}
F_0=-0.0049488147(2).
\end{equation}
The other parameters $A_2,\cdots,A_8$ are fitted and listed in
Tab. \ref{fitfen}.  For the infinitely long strip, $A_2
\rho^{-3/2},A_3 \rho^{-2}$ and $A_4\rho^{-5/2}$, at the limit
$\rho \to \infty$, correspond to the coefficients  of $N^{-3},
N^{-4}$ and $N^{-5}$ terms in the finite size free energy
expansion respectively, which have been obtained as $-0.04616(2),
0.024(1), 0.69(6)$, by using numerical transfer matrix techniques
\cite{queiroz}.
Simple extrapolations of the fitted values of
$A_2,A_3,A_4$ 
show that the magnitude of
$A_2,A_3$ agree with the transfer matrix results, but $A_4$ does
not. 
\begin{ruledtabular}
\begin{table*}[htbp]
 \caption{ The fitted parameters of Eq.  (\ref{freeenergy}) for the critical free energy. }
\begin{tabular}{cccccc}
 parameter   & $\rho=1$ & $\rho=2$ & $\rho=4$  & $\rho=8$ & $\rho=16$\\
\hline
$A_1$ & $0.060969202833(3)$ & $0.08263005222(2)$
 & $0.170207175502(4)$  & $0.388684864512(8)$ & $0.86896194132(2)$ \\
$A_2 $ & $0.0883883476(2)$ & $0.0595556310(1)$
 & $-0.1009034881(4)$  & $-0.666287857(1)$ & $-2.423248445(3)$ \\
$A_3 $ & $-0.0175362651(2)$ & $0.04067435(2)$
 & $0.41842401(7)$  & $2.1854954(3)$ & $9.76558(1)$ \\
$A_4 $ & $-0.02405666(2)$ & $-0.174428(2)$
 & $-1.290133(9)$  & $-8.15503(5)$ & $-48.556(3)$ \\
$A_5 $ & $0.066893(8)$ & $0.4351(1)$
 & $3.9657(8)$  & $33.676(6)$ & $277.21(5)$ \\
$A_6 $ & $-0.16558(2)$ & $-1.147(4)$
 & $-13.85(5)$  & $-161.7(5)$ & $-1857(6)$ \\
$A_7 $ & $0.3851(4)$ & $3.22(8)$
 & $53(1)$  & $86(2)0$ & $137(4)00$ \\
$A_8 $ & $-0.703(3)$ & $-7.6(8)$
 & $-17(2)0$  & $-387(4)0$ & $-8(1)0000$ \\
\end{tabular}
\label{fitfen}
\end{table*}
\end{ruledtabular}
We have tried other forms of formula to fit the critical free
energy data. For example, we added the terms $\ln S/S^{3/2}, \ln
S/S^2$ in the fitting formula and found that the corresponding
coefficients are extremely small (less than $10^{-7}$).  We
concluded that the logarithmic correction only appears in the
corner term $\ln S/S$. We  note here that, in the asymptotic
expansion of the free energy for the Ising model with periodic
boundary conditions \cite{izmailian2002a}, with Braskamp-Kunz
boundary conditions \cite{izmailian2002b} and with helical
boundary conditions \cite{izmailian2007}, only integer powers of
$S$ appear.

{\it Critical internal energy}.  We fit the data of critical
internal energy  with the formula given by Eq.
(\ref{internalenergy}) with $k$ from 1 to 4. The bulk value $U_{\infty}$ is known to be
$\sqrt{2}\approx 1.41421356237$ \cite{onsager}. Our fit of
$U_{\infty}$ is $1.4142135624$ for the five aspect ratios.
\begin{ruledtabular}
\begin{table*}[hbtp]
\caption{The fitted parameters of Eq. (\ref{internalenergy}) for the
critical internal energy per spin.}
\begin{tabular}{cccccc}
 parameter   & $\rho=1$ & $\rho=2$ & $\rho=4$  & $\rho=8$ & $\rho=16$\\
\hline $U_{surf} $ & $-0.63661981(1)$ & $-0.63661983(3)$
 & $-0.63661985(4)$  & $-0.63661987(5)$ & $-0.63661988(5)$ \\
$B_1 $ & $-0.1213621(2)$ & $-0.2586182(5)$
 & $-0.6453965(8)$  & $-1.266482(1)$ & $-2.151535(2)$ \\
$U_{corn} $ & $-0.450170(3)$ & $-0.45018(1)$
 & $-0.45022(2)$  & $-0.45028(4)$ & $-0.45042(9)$ \\
$B_2 $ & $-0.98604(2)$ & $-1.1278(9)$
 & $-1.7138(2)$  & $-3.2016(4)$ & $-6.4893(9)$ \\
$B_3 $ & $-0.2969(2)$ & $-0.101(1)$
 & $0.833(3)$  & $3.943(8)$ & $13.40(2)$ \\
 $B_4 $ & $-0.040(2)$ & $-0.41(1)$
 & $-2.40(4)$  & $-11.2(2)$ & $-47.7(7)$ \\
\end{tabular}
\label{fitien}
\end{table*}
\end{ruledtabular}
The leading correction should be caused by the edges. In the exact
result of Au-Yang and Fisher \cite{fisher} on the strip with two
free edges, the edges' correction is given by $-\frac{2}{\pi}\ln
N/N$. We conjecture that, on the rectangle with four free edges,
the edge correction is given by $-\frac{2}{\pi}(M\ln N+N\ln
M)/(MN)$ with the coefficient
$U_{surf}(\rho)=-\frac{2}{\pi}=-0.636619773\dots$. 
This conjecture is proved by the fitted $U_{surf}$ for all $\rho$
($U_{surf}=-0.6366198...$), which agree with the predicted value
in the accuracy $10^{-7}$.

The term $B_1/S^{1/2}$ is in fact scaled as $1/N$. In the infinity
long strip limit, the coefficient before $1/N$ is known as
$\frac{2}{\pi}(\frac{7}{2}\ln 2+\gamma-\frac{\pi}{4}-\ln
\pi)\approx 0.683158$ \cite{fisher}. As we can see in Tab.
\ref{fitien}, $B_1/\sqrt{\rho}$ indeed approaches this limit as
$\rho$ increases.

Following the convention in the critical free energy, we write the
coefficient of $(\ln S)/S$ as $U_{corn}$. We found
\begin{equation}
U_{corn}(\rho)\approx -0.45025(3), \label{eq:u-corn}
\end{equation}
which is independent of aspect ratio $\rho$.
Apparently, 
this contribution becomes zero in the limit of infinitely long
strip, in which there is no corner. From this point, it is
rational to call this term corner's correction.

The other parameters $B_1, B_2, B_3, B_4$ are fitted and listed in
Tab. \ref{fitien}. Again we have tried other forms of formula to
fit the critical internal energy. The terms $\ln S/S^{3/2}, \ln
S/S^2$ are excluded considering the coefficients are extremely
small. Moreover the standard deviations of the fits with these
terms are much larger than those without them.

{\it Critical specific heat}. The data of the critical
specific heat are fitted using the formula given by Eq. (\ref{specificheat})
with $k$ from 1 to 4. The leading term $A_0\ln N$ is known from Onsager's exact result
\cite{onsager}, which reads
$A_0=\frac{2}{\pi}[\ln(1+\sqrt{2})]^2\approx 0.494538589$. Our
fitting gives $A_0\approx 0.49453858$. The other fitted
parameters are shown in Tab. \ref{fitsh}.
\begin{ruledtabular}
\begin{table*}[hbtp]
\caption{The fitted parameters of Eq.
(\ref{specificheat}) for the critical specific heat.}
\begin{tabular}{cccccc}
 parameter   & $\rho=1$ & $\rho=2$ & $\rho=4$  & $\rho=8$ & $\rho=16$\\
\hline
 $c_0 $ & $-0.57078599(3)$ & $-0.44276294(2)$ & $-0.37766115(2)$ & $-0.34510742(2)$ & $-0.32883055(2)$ \\
$c_{surf} $ & $0.524516(2)$ & $0.524533(2)$ & $0.524531(2)$  & $0.524530(3)$ & $0.524529(3)$ \\
$D_1 $ & $-0.34928(3)$ & $-0.29524(1)$ & $-0.17826(1)$  & $-0.05986(2)$ & $0.03117(2)$ \\
$c_{corn} $ & $0.3683(7)$ & $0.3704(3)$ & $0.3696(7)$  & $0.368(1)$ & $0.365(3)$ \\
$D_2 $ & $1.144(5)$ & $1.293(2)$ & $1.914(5)$  & $3.42(1)$ & $6.67(3)$ \\
$D_3 $ & $0.02(4)$ & $-0.13(2)$ & $-1.25(7)$  & $-4.8(2)$ & $-15.4(6)$ \\
$D_4 $ & $0.8(3)$ & $0.7(2)$ & $3.6(8)$  & $16(3)$ & $7(1)0$ \\
\end{tabular}
\label{fitsh}
\end{table*}
\end{ruledtabular}
The coefficient $c_0$ increases with the aspect ratio. For the
strip case it is known that $-c_0(\rho=\infty)=(\frac{7}{2}\ln
2+\gamma-14\xi(3)/\pi^2-\frac{\pi}{4}-\ln \pi)\approx 0.3125538$
\cite{fisher}. $c_0$ approaches this limit as $\rho \to \infty$
obviously, see Tab. \ref{fitsh}. However we have not obtained an
analytical expression for the dependence of $c_0$ on $\rho$.

The term $(M\ln N+N\ln M)/S$ is the next order correction. Its
coefficient $c_{surf}$ is independent of $\rho$, and its average
value over $\rho$ is $c_{surf}=0.524529(3)$.
Note that this term is absent in the torus case \cite{fisher1969}
and not mentioned in the long strip case \cite{fisher}, but
exists in the cylinder case with Brascamp-Kunz
boundary conditions \cite{Janke,izmailian2002b}.

The corner term $c_{corn}$ also seems independent of aspect ratio $\rho$ and its
average value is about
\begin{equation}
c_{corn}=0.368(1).
\end{equation}

We have also tried many other forms of formula to fit the critical
specific heat data. For example, we added the terms $(\ln S)^3/S,
(\ln S)^2 /S $ in the fitting formula and found that their
coefficients are extremely small.

{\it Conclusion}. Using the BP algorithm, we studied the Ising
model on a rectangle of size $M \times N$ with free boundaries.
For five aspect ratios $\rho=1,2,4,8,16$, the critical free
energy, internal energy and specific heat were calculated. We
brought numerical evidence that the correction term $f_{corn}$
stems from the corners of the rectangle is indeed universal and is
proportional to the central charge $c$. We also found that the
terms $U_{corn}$ and $c_{corn}$ are independent from the aspect
ratio $\rho$. In order to check whether or not the terms
$U_{corn}$ and $c_{corn}$ are universal quantities, it is useful
to extend our study to the Ising model on other types of lattices,
e.g., the triangular, honeycomb, Kagome lattices. In such studies,
the BP algorithm which is suitable for any planar network of Ising
spins with arbitrary bond strengths \cite{loh1,loh2} is still a
powerful tool. In addition, we can enhance the accuracy of  the
internal energy and specific heat by using an extended BP
algorithm to calculate the internal energy without the using of
differentiation\cite{loh2}. Moreover the finite size effects of
the correlation can be investigated by using the site propagation
algorithm \cite{wu}. As we have shown, the sharp corners induce
remarkable effects in critical region not only on the free energy,
but also on the internal energy and the specific heat. It is
expected that the sharp corners can induce remarkable effects on
other properties of finite size systems in critical regime, for
example, thermal conductivity, electric conductivity, etc. All of
these effects should be observable in experiments.

{\it acknowledgment}
This work is supported by the National Science Foundation of China (NSFC)
under Grant No. 11175018.

\end{document}